\documentclass{aa}
\usepackage{graphicx}
\graphicspath{{./}{figures/}}
\usepackage{natbib}
\usepackage{amsmath}
\usepackage{verbatim}
\usepackage{xcolor}
\usepackage{txfonts}
\usepackage[colorlinks=true]{hyperref}
\hypersetup{linkcolor=blue,citecolor=blue,filecolor=blue,urlcolor=blue}
\usepackage{ulem,lipsum}

\begin{document} 

\title{H$\beta$ line shape and radius-luminosity relation in 2.5D FRADO}

\titlerunning{Characteristics of H$\beta$ line in 2.5D FRADO}

   \author{M. H. Naddaf
          \inst{1}\fnmsep\thanks{mh.naddaf@uliege.be}
          \and
          M. L. Martinez-Aldama
          \inst{2,3,4}
          \and
          D. Hutsem\'ekers
          \inst{1}
          \and
          D. Savic
          \inst{5,1}
          \and
          B. Czerny
          \inst{6}}

        \institute{Institut d'Astrophysique et de Géophysique, Université de Liège, Allée du six août 19c, B-4000 Liège (Sart-Tilman), Belgium
         \and
        Astronomy Department, Universidad de Concepción, Barrio Universitario S/N, Concepción 4030000, Chile
        \and
        Millennium Nucleus on Transversal Research and Technology to Explore Supermassive Black Holes (TITANs), Concepción, Chile
        \and
        Millennium Institute of Astrophysics (MAS), Nuncio Monseñor Sótero Sanz 100, Providencia, Santiago, Chile
         \and
        Astronomical Observatory Belgrade, Volgina 7, 11060 Belgrade, Serbia
         \and
        Center for Theoretical Physics, Polish Academy of Sciences, Al. Lotnik\'ow 32/46, 02-668 Warsaw, Poland
             }

\date{Submitted: May 2025}

 
  \abstract
   {Galaxies with active galactic nuclei (AGNs) exhibit broad emission lines as a key spectral feature. The shape of the emission-line profiles depends on the complex dynamics of discrete clouds within a spatially extended region known as the broad line region (BLR). The distribution of cloud positions within the BLR (or the de facto geometry of BLR) is directly linked to measurements of BLR time lags.}
   {In this work, we convolved a large grid of physically based simulations of cloud distributions in BLR with photon-flux weighted emissivity of BLR clouds to investigate the generic shape of spectral line profiles. More importantly, we were able to extract the time-delay histograms of corresponding models to calculate the size of BLR.}
   {Our physical model is based on the assumption that the clouds are launched by the radiation pressure acting on dust in the atmosphere of the outer disk. It has very few global parameters. The model is appropriate for the low-ionization part of the BLR, as seen in earlier model tests. It uses a non-hydrodynamical single-cloud approach to the BLR dynamics, which enables us to simulate the distribution of positions and velocities of the clouds.}
   {We found that the line profile width broadens with increasing black hole mass (or with viewing angle) and narrows with increasing accretion rate. The blue wing of the emission line profiles becomes more pronounced with increasing black hole mass and accretion rate, consistent with the formation and intensification of an outflow structure.  We also found that the peak time delays offer a better representation of the observational trend and the scatter in the radius-luminosity (R-L) relation, compared to averaged delay values.‌}
   {}


\maketitle


\section{Introduction}

Broad emission lines in the optical-UV spectra of active galactic nuclei (AGNs) are distinct features characterized by their significant width of several 1000 km/s. These lines are generated in the BLR, a spatially extended region near the central supermassive black hole in an AGN \citep{seyfert1943, schmidt1963}. The BLR consists of dense, rapidly moving clouds of gas that are photoionized by the intense optical-UV continuum radiation emitted from the accretion disk surrounding the black hole. As a result, these clouds emit spectral lines that are Doppler-broadened due to their high velocities, and they can even become relativistically broadened if they get close enough to the central black hole, leading to the observed broad emission line profiles in the spectra of AGNs \citep{baldwin1997,krolik_book1999,sulentic2000,bianchi2019}.

The results of recent studies indicate that BLRs are universally present for accreting supermassive black holes. This is true for  both type I AGNs, where the BLR is not obscured by the torus \citep{antonucci1993}, and type II AGNs, where the scattered polarized light can reveal the broad lines \citep{antonucci1985}. The only exception is when the luminosity or, correspondingly, the accretion rate, is very low \citep{sabra2003, genzel2010, eckart2017}, indicating that the presence of BLR depends on the accretion rate \citep{czerny2004a,elitzur2009, du2015, DuPu2018, naddaf2021, naddaf2022}. Although some sources with low accretion rates have been reported to contain BLR \citep{bianchi2019}, in most low-luminosity sources, a hot optically thin flow replaces the inner standard disk and the BLR gets reduced to a narrow ring \citep[see e.g.,][for a review]{eracleous2009}; thus, the emission seems to come directly from the illuminated disk surface. The details of the disappearance of the cold disk by the BLR remain poorly understood \citep[e.g.,][]{liu1999, rozanska2000, nicastro2003, czerny2004a, liu2022} and may depend on the presence of the magnetic field and hot plasma surrounding the nucleus. In extreme cases, such as that of Sgr A*, there is no cold disk or BLR present. In brief, Sgr A* is a very dim source with mass accretion rate of $10^{-9}$ to $10^{-7} \ M_\odot/\text{yr}$ and Eddington ratio of ${L}/{L_{\rm Edd}} \sim 10^{-9}$ to $10^{-8}$ \citep{aitken2000, bower2003, marrone2007, yoon2020}; this can be compared to the typical range of Eddington ratios in AGNs from $10^{-5}$ up to a few.

Overall, BLRs are  spatially unresolved, except in recent interferometric measurements reported for seven different sources \citep{GRAVITY3C273_2018, GRAVITY3783_2021, GRAVITYIRAS_2020, GRAVITY_4_objects_2024}. However, the variability of the central source and the delayed response of BLR to variations in the continuum allows for the measurement of the characteristic distance from BLR to the central black hole. Given the constancy of the speed of light, $c$, the distance is $ R_{\rm BLR} = c \tau$ , where $ \tau $ represents the time lag between the continuum and a selected BLR emission line,  measured using the reverberation mapping (RM) technique \citep{blandford1982, peterson1993,peterson2014}. The technique involves monitoring the AGN continuum (typically measured at 5100 \AA) and broad emission lines (most notably, H$\beta$) over extended periods. Combining the BLR radius with the broad emission-line velocity width yields the virial mass enclosed within the BLR, which is dominated by the black hole \citep{peterson1998, kaspi2000}. The validity of reverberation masses is supported by other means; for example, by the consistency of reverberation-measured masses with other dynamical mass measurement methods \citep{davies2006, onken2007, hicks2008}.

Optical reverberation mapping studies using the H$\beta$ line have revealed a power-law relation of  $R_{\rm BLR} = L_{\lambda}^{\alpha}$ between the size of the BLR and the monochromatic luminosity of AGNs at 5100 \AA~\citep{kaspi2000, peterson2004}. This is known as the R-L relation, where the $\alpha$ value is typically close to 0.5 for H$\beta$ emission lines. Despite its success, the R-L relation exhibits scatter, which has increased with the accumulation of more sources \citep{du2015}. This scatter is often linked to accretion rate-dependent changes in the BLR structure, with highly accreting sources showing the largest departures from the nominal R-L relation \citep{du2015, DuPu2018}. Several studies have explored the origin of this scatter \citep[e.g.,][]{wang_shielding2014, czerny2015, czerny2017, woo2024}, however,  corrections based on the Eddington ratio and dimensionless accretion rate to reduce the scatter yielded a tighter relation \citep{DuPu2018, martinez2019}, followed by significant progress achieved by incorporating the relative Fe II strength, which correlates with accretion-rate intensity \citep{Dupu2019}.

Despite considerable efforts, scatter remains an open question, with  investigations underway from both theoretical and observational perspectives. The scatter is not confined to the R-L relation derived from reverberation-measured H$\beta$ time-delays, but also extends to other emission lines such as Mg II and C IV. This motivates continued investigation of the underlying causes of the scatter in different emission lines. 

Another active area involves using BLR time delays for cosmology. The idea was outlined in several recent publications \citep{watson2011, czerny2013, martinez2019} and some applications have been made \citep{khadka2021,cao2022,cao2024}; although we note that these have relied on a parametric approach to the R-L relation. Without external calibration, this approach can bring the results from the curvature of the Hubble diagram, constraining the $\Omega_m$ but not the Hubble constant. However, with a physically motivated model predicting the position of the BLR for a given source intrinsic (absolute) luminosity, it would be also possible to measure the Hubble constant by combining the time delay observations with the observed source flux. 

In this paper, we investigate BLR properties from a theoretical perspective, without focusing on specific observational datasets.We present simulations of BLR geometry and dynamics using our 2.5D  Failed Radiatively Accelerated Dusty Outflow (FRADO) model, which specifically accounts for the low-ionized part of the BLR \citep{naddaf2021}. From our simulations, we were able to calculate theoretical transfer functions, map the BLR, and determine a representative single-value delay to compare these parameters across a distribution of global parameters with the observational trends in the R-L relation. 

The structure of the paper is as follows. In Section \ref{sec:BLRmodelling}, we present the modeling of the BLR. Section \ref{sec:profiles} details the computation of emission line profiles based on our model, followed by Section \ref{sec:blrsize}, where we address the calculation of time-delay maps and extraction of R-L relation from our simulations. In Section \ref{sec:discussion}, we discuss our results. Finally, we present our conclusions in Section \ref{sec:conclusion}.

\section{Modeling the BLR}\label{sec:BLRmodelling}

To model the BLR, we used the 2.5D FRADO model, which is based on a physically motivated dust-driven outflow from the disk surface in the region where the disk surface temperature is lower than the sublimation temperature. This mechanism sets the BLR location and governs cloud dynamics. Unlike parametric BLR models \citep[e.g.,][]{pancoast2011, pancoast2012, li2013}, it does not rely on arbitrary parametrization of cloud positions. It adopts a non-hydro, single-cloud dynamical approach to the BLR dynamics.

\subsection{The 2.5D FRADO}\label{sec:FRADO}

The FRADO model describes the formation of the BLR in an AGN. This model is based on the concept that radiation pressure acts on dust in  regions of the accretion disk where the temperature of the disk atmosphere is below the dust sublimation threshold. According to FRADO, this mechanism is responsible for the uplift of material that contributes to the emission of low-ionization lines (LIL). The model was initially proposed by \citet{czerny2011} and  later extended to the 2.5 dimension by \citep{naddaf2021}. The model posits that radiative pressure on dust lifts clouds from the disk surface, while preserving the angular momentum derived from the Keplerian motion of the disk. Once lifted, these clouds are more strongly illuminated by the inner disk regions. If a cloud becomes too hot, the dust evaporates and the cloud continues its motion along a ballistic orbit. 

The global parameters of the model are the black hole mass, the Eddington rate, and the metallicity. Under the conditions of a lower black hole mass, lower Eddington ratio, and lower metallicity, the clouds can only form a failed wind. Conversely, in scenarios of higher black hole mass, higher Eddington ratio, and higher metallicity, a fraction of the clouds might form an escaping wind. Early tests of the model showed consistent results with observed trends \citep{czerny2015,naddaf2022,naddaf2024, naddaf2024Univ}, including applications to broad absorption line quasars \citep{naddaf2023}. In this work, we test the model more systematically across a broad parameter space.

Given the global source parameters, we can compute the 3D locations and velocities of statistically representative BLR clouds. The method for computing single-cloud dynamics is detailed in \citet{naddaf2021}. We can then apply an emissivity law to the cloud distributions to model LILs, such as H$\beta$, Mg II, or Fe II.

\subsection{Simulation setup and the grid of models}
\label{sect:grid}

To perform a comprehensive study, we assumed a wide and dense grid of black hole masses ($\log M_{\bullet}$ from 6 to 10 in solar units), and accretion rates ($\log \dot{m}$ from -2 to 0 in Eddington units), both sampled in 0.2 dex intervals. This grid spans the typical parameter range of quasars in the SDSS catalog \citep{shen2011}. The dust sublimation temperature is fixed at 1500 K \citep{barvainis1987}, and metallicity is set at five times solar, corresponding to a dust-to-gas mass ratio of 0.025 \citep{baskin2018, naddaf2022}. The adopted range of the viewing angles covers the range from 10$^\circ$ to 60$^\circ$, with the step of 5$^\circ$. Larger inclinations (60$^\circ$) have been excluded, as the BLR becomes obscured by the dusty torus, and such sources appear as type 2 AGNs, with only narrow lines visible in direct light \citep[see ][for a general review]{krolik_book1999}. 

\section{LIL BLR line profiles}\label{sec:profiles}

\subsection{Line profile modeling}

Given a distribution of BLR clouds along with their velocity field and local emissivity, we can find the shape of the emission line profiles. As the LIL-emitting BLR lies at radii larger than a few 100 $r_g$ (regardless of the black hole mass), the gravitational redshift can be neglected and only the Doppler effect contributes to line broadening. Previously, we assumed that the BLR simply follows a single-emissivity function weighted with the vertical position of a given cloud in the ensemble of clouds forming the BLR \citep{naddaf2022}. We can now refine this approach by convolving the line emissivity with the location-dependent flux of ionizing photons, $\phi_{H}(r)$.

\begin{figure}
    \centering
    \includegraphics[width=7cm]{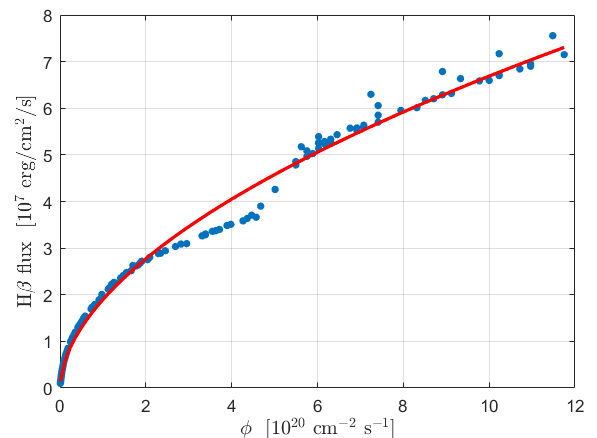}
    \caption{Results from a CLOUDY simulation (Ashwani Pandey; private communication) showing the H$\beta$ line emissivity as a function of the incident photon flux $\phi$ (computed from the 2.5D FRADO model) reaching a BLR cloud. The blue points represent simulation outputs, while the red curve shows a logarithmic fit, highlighting a clear nonlinear relationship. \label{fig:emissivity}}
\end{figure}

Clouds are also assumed to be optically thin, so that each is uniformly illuminated in all directions (i.e., spherically symmetric illumination; no self-occultation or cloud shadowing, known as the moon effect; \citealt{rosborough2024}). Each cloud then reprocesses the absorbed radiation into isotropic line emission.
Preliminary BLR simulations using the 2.5D FRADO model were presented in \citet{naddaf2021, naddaf2022}. Here, we have carried out a systematic survey aimed  at also modeling the time delay.
We adopted a universal conversion factor between local irradiating flux and line luminosity for all clouds. The H$\beta$ line emissivity was assumed to scale approximately with the logarithm of the local ionizing photon flux, which is a function of the cloud's location \citep{gilbert2003, korista2004}.
At the photoionization equilibrium, the flux of hydrogen recombination lines, such as H$\beta$, generally scales with the number of ionization per unit volume, which  is itself tied to the incident ionizing photon flux $\phi$. At low $\phi$, this relation is approximately linear; however, at higher photon fluxes, due to saturation effects in ionization and optical depth, the scaling becomes sub-linear (more slowly than linearly). This is consistent with expectations in ionization-bounded or partially matter-bounded regimes. In such cases, increased ionizing photon flux leads to diminishing returns in line emission due to saturation and thermal balance effects.
Photoionization simulations using CLOUDY \citep{ferland1996, moloney2014, kriss2019, temple2021, CLOUDY2023, pandey2023} confirm that a logarithmic dependence of H$\beta$ emissivity on $\phi$ provides an excellent empirical fit.  Figure \ref{fig:emissivity} shows the H$\beta$ line emissivity (flux) as a function of the photon flux $\phi$, for $\phi > \sim 10^{17}$ photons cm$^{-2}$ s$^{-1}$ (Ashwani Pandey; private communications). The photon flux and the ionizing continuum spectral energy distribution (SED), with respect to the location of each cloud, were directly computed from the 2.5 FRADO. These results were then input into CLOUDY, along with fixed values for the gas density, $n_{\rm H}$ of $10^{11}$ cm$^{-3}$, and column density, $N_{\rm H}$ of $10^{23}$ cm$^{-2}$. The results confirm a non-linear H$\beta$ response to photon flux, which is well-approximated by a logarithmic trend \citep[see also Figure 2 in][for NGC 5548]{gilbert2003}.

\subsection{Results}

\subsubsection{Dependence of line shapes on global parameters}

\begin{figure*}[htbp]
        \centering
        \includegraphics[scale=0.45]{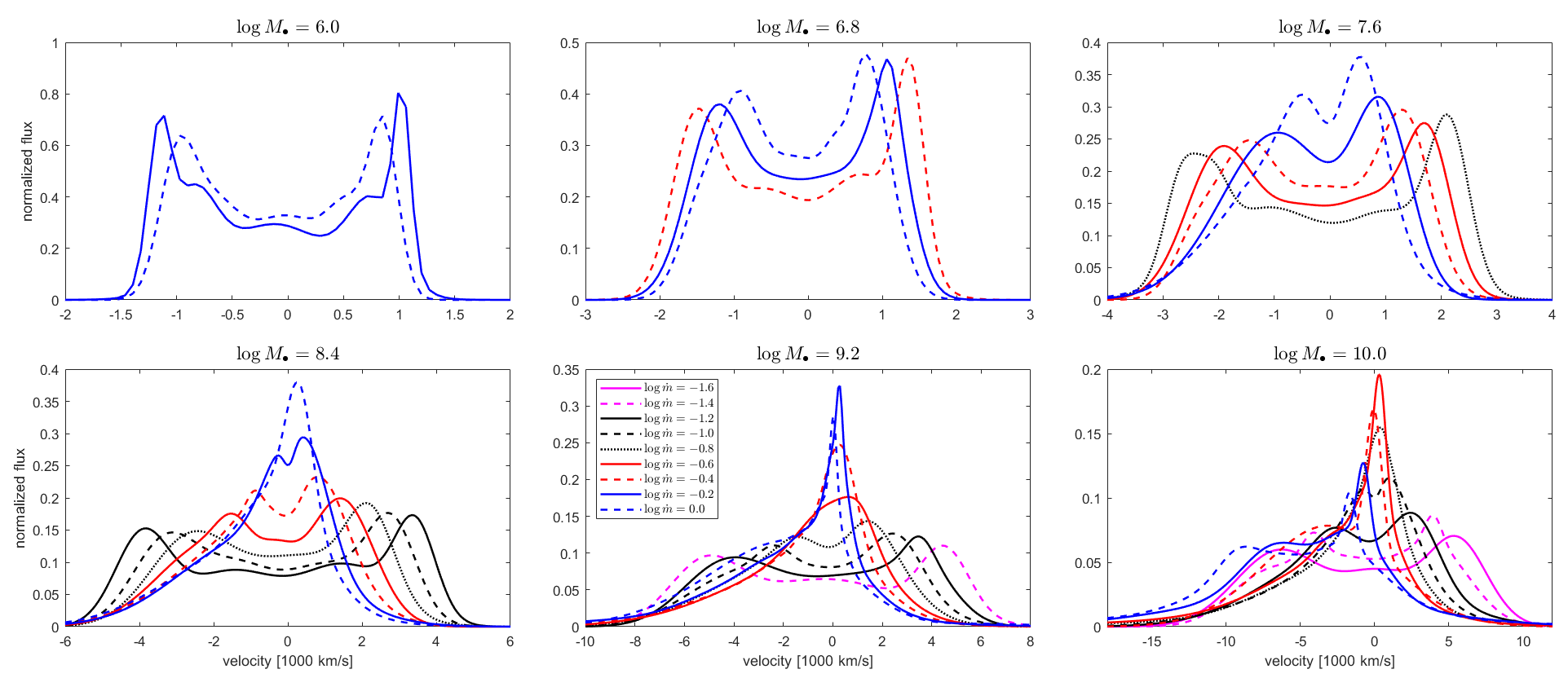}
        \caption{Predicted H$\beta$ line profiles grouped by black hole mass, assuming a fixed viewing angle of 39$^\circ$. Different curves within each panel correspond to representative Eddington ratios (see text for details). Note: the horizontal axis scale varies between panels. All profiles are normalized to the unit total flux.}
        \label{fig:line_mass}
\end{figure*}

\begin{figure*}[htbp]
        \centering
        \includegraphics[scale=0.45]{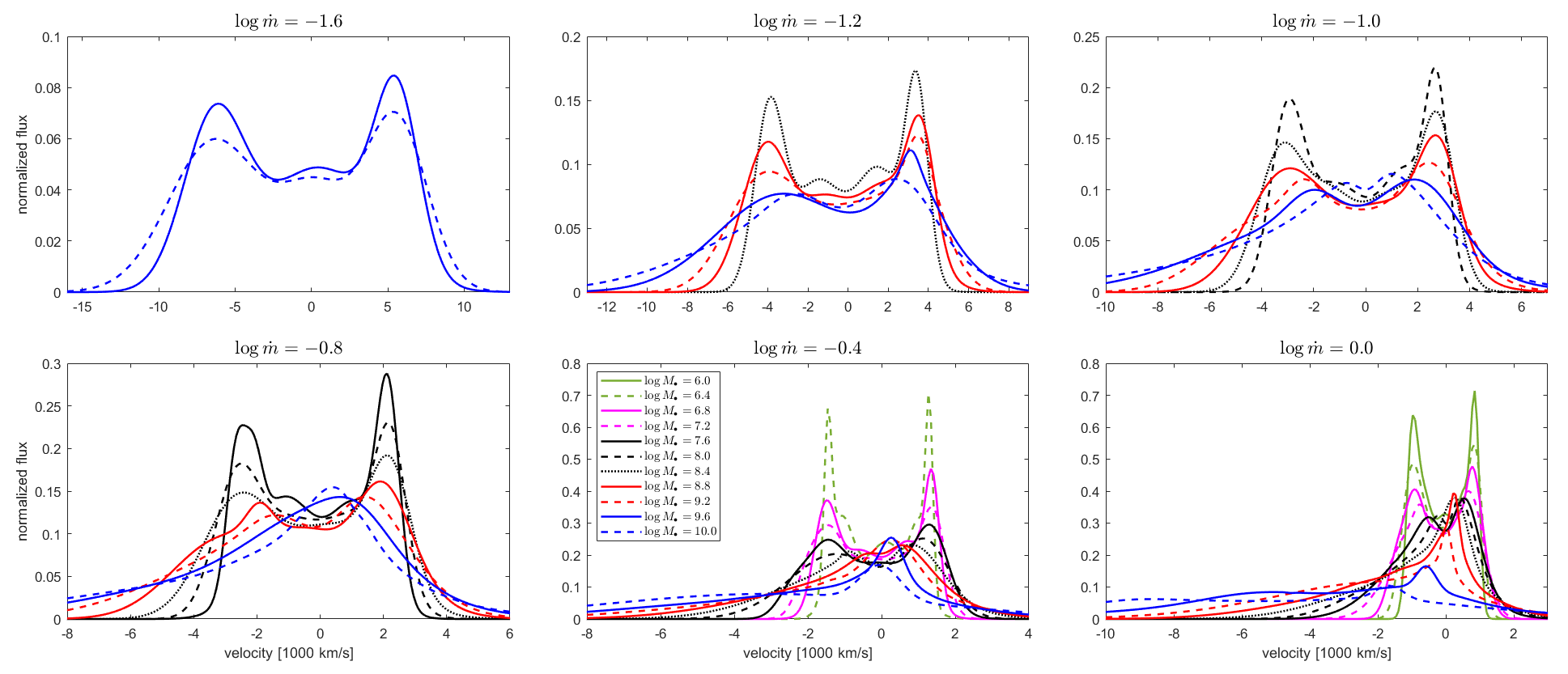}
        \caption{Predicted H$\beta$ line profiles grouped by Eddington ratio, assuming a fixed viewing angle of 39$^\circ$. Different curves within each panel correspond to representative black hole mass (see text for details). Note: the horizontal axis scale varies between panels. All profiles are normalized to the unit total flux.}
        \label{fig:line_mdot}
\end{figure*}

\begin{figure*}[htbp]
        \centering
        \includegraphics[scale=0.5]{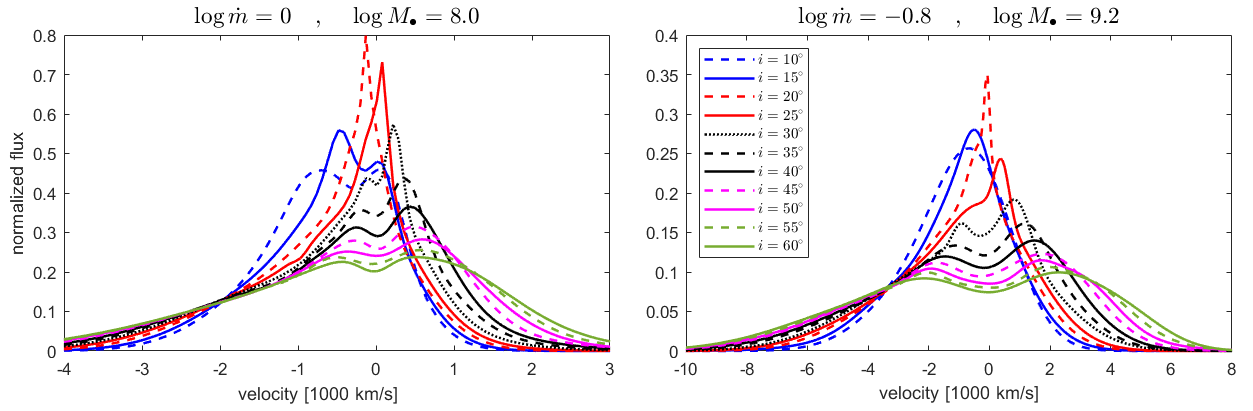}
        \caption{Dependence of H$\beta$ line profile shape on viewing angle. Results are shown for two representative models from our simulation grid. The left panel corresponds to a canonical AGN model and the right panel represents the mean quasar. All profiles are normalized to unit total flux.}
        \label{fig:line_view}
\end{figure*}

Figures \ref{fig:line_mass}, \ref{fig:line_mdot}, and \ref{fig:line_view} present LIL emission line profiles  from the full model grid defined in Section~\ref{sect:grid}. The figures are presented in velocity space with positive velocities defined as blueshift. Since the adopted scaling with incident flux was calibrated for the H$\beta$ line, the following results primarily pertain to H$\beta$.

Figure \ref{fig:line_mass} shows the line profiles at a representative viewing angle of 39$^\circ$ \citep{lawrence2010} for varying black hole masses and Eddington ratios, as indicated in the legend. For low black hole masses, only high Eddington ratios are shown, since at low accretion rates even the failed wind is not launched. In such cases, lines may form directly in the disk, but the inner BLR radius is not well defined within the FRADO model. This issue is discussed further in Section~\ref{sec:discussion}. As the black hole mass increases, the profiles become broader, consistent with deeper gravitational potential wells and higher orbital velocities in the BLR. At lower masses, the line profiles show a pronounced double-peaked structure, indicative of a disk-dominated BLR geometry \citep[e.g.,][]{eracleous94}. This occurs at all Eddington ratios where a FRADO solution exists. However, this is inconsistent with observations, which show single-peaked profiles for high Eddington ratios and small black hole masses \citep[e.g.,][]{moran1996, laor1997, veron_cetty2001, cracco2016}. This suggests that vertical velocities may be underestimated by FRADO at low black hole masses. The double-peaked feature diminishes for higher masses, where high accretion rates produce single-peaked, broad profiles, while double peaks remain present in the low accretion regime. This trend is well aligned with observations.

Figure \ref{fig:line_mdot} shows the dependence of the H$\beta$ line profile on Eddington ratio, keeping the black hole mass constant in each panel, for values between $10^6$ and $10^{10} M_\odot$ and at a representative viewing angle of 39$^\circ$. As the accretion rate increases, the H$\beta$ line profiles generally become narrower, with the full width at half maximum (FWHM) decreasing from $\sim 10^4$ km/s to $\sim 10^3$ km/s. However, the line also develops a blueshifted asymmetry as the accretion rate rises. At very high accretion rates, close to the Eddington limit and for high masses, the line broadens again, as most of the emission originates in an outflow.

At lower accretion rates, radiation pressure is weak or negligible, and outflows are suppressed or absent. The BLR forms closer to the black hole, where gas remains gravitationally bound and moves at higher orbital velocities, producing broader, double-peaked profiles. Conversely, at high accretion rates, radiation pressure and outflows dominate, driving winds and potentially increasing turbulence in the BLR. This leads to more complex velocity fields and blueshifted asymmetric profiles, particularly near the Eddington limit. These results are consistent with observational studies, which show that AGNs with high accretion rates tend to have blueshifted H$\beta$ profiles due to radiation-driven winds, while low accretion rate AGNs exhibit more symmetric lines \citep{negrete2018, marinello2020}.

Strong outflow signatures are more evident in high-ionization lines (HILs) such as C IV $\lambda1549$, which originate closer to the continuum source and are strongly influenced by radiation forces \citep[e.g.,][]{coatman2017, mlma2018, temple2021}. However, these lines arise in the HIL region where radiation line-driving, rather than dust-driving, dominates; thus, they lie beyond the scope of our model.

Figure \ref{fig:line_view} shows the H$\beta$ profiles for a canonical AGN model with a black-hole mass of $10^{8} M_{\odot}$ accreting at the Eddington rate (left panel) and for a mean quasar model, with a higher black hole mass and lower Eddington ratio, drawn from \citet{shen2011} (see the right panel). Profiles are plotted for viewing angles of 10$^\circ$ to 60$^\circ$, measured from the symmetry axis. The dominant trend is line broadening with increasing inclination, driven by greater projection of the BLR’s rotational velocity along the line of sight. The line core is generally blueshifted, while the red wing remains highly sensitive to viewing angle, reflecting the underlying outflow geometry.

\subsubsection{Parameters of line shape}

\begin{figure*}[ht!]
        \centering
        \includegraphics[scale=0.7]{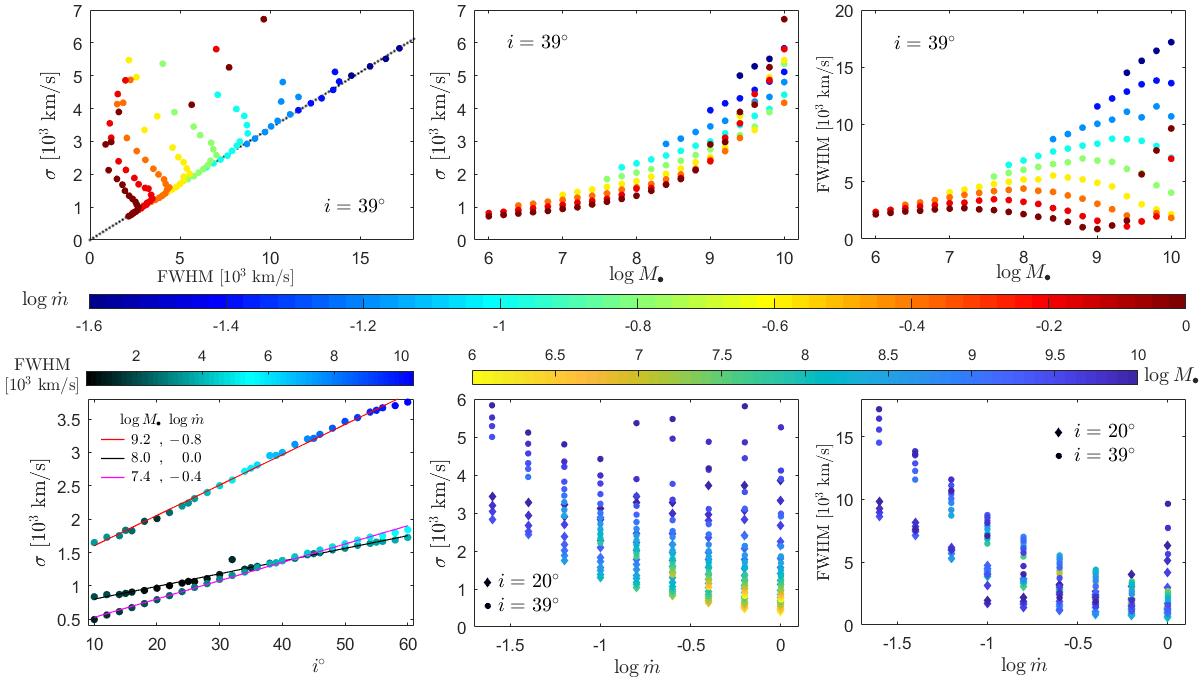}
        \caption{Relations between $\sigma$ and FWHM, and the predicted dependence on the black hole mass, accretion rate, and viewing angle. The black dotted line stands for an arbitrary linear fit, $\sigma = 0.34$ FWHM, to the trend of the underneath of data points in top-left panel. The solid lines in the bottom-left panel are also linear fits to the data points.}
        \label{fig:sigma_FWHM}
\end{figure*}
\begin{figure*}
        \centering
        \includegraphics[scale=0.45]{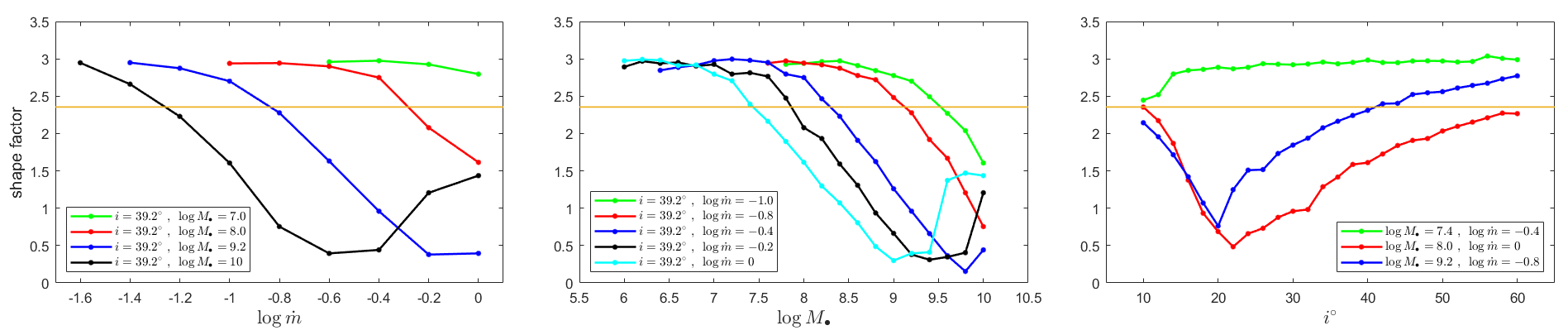}
        \caption{Behavior of shape factor, the ratio of $\rm FWHM$ to $\sigma$, with respect to the accretion rate, and black hole mass of the source, and also versus the observer's viewing angle. The orange line marks the value expected for a Gaussian line shape and separates Gaussian-like profiles (upper part of the plots) from Lorentzian-like profiles (lower part of the plots).}
        \label{fig:line_shapefactor}
\end{figure*}

Line shapes are most frequently characterized by either the FWHM or the standard deviation ($\sigma$). These two standard parametrizations are not equivalent and are typically associated with Gaussian or Lorentzian shapes, as discussed by \citet{collin2006}. Since the model provides the full line shape, we can examine how these shapes map onto these standard parameters.

Figure \ref{fig:sigma_FWHM} shows the relationship between the standard deviation ($\sigma$) and the FWHM of emission lines, as well as other global parameters, for a representative fixed viewing angle. There is a positive correlation between $\sigma$ and FWHM, indicating that broader lines generally exhibit higher $\sigma$ values. Higher Eddington ratio models tend to show smaller $\sigma$ and FWHM, while lower accretion rates yield broader profiles with larger $\sigma$. A clear increasing trend of $\sigma$ with black hole mass is observed, while FWHM shows a more complex dependence; however, we note that  both parameters decrease with increasing Eddington ratio.

For a fixed Eddington ratio of 0.3, often considered a transitional value between low and high accretion \citep[e.g.,][]{laor1989, campitiello2019}, the model yields $\sigma \sim 1000$ km s$^{-1}$ for a black hole mass of $10^7M_{\bullet}$ and about twice that for $10^9M_{\bullet}$. This aligns well with the transition between Seyfert 1 galaxies and Narrow-Line Seyfert 1 galaxies at FWHM $\sim 2000$ km s$^{-1}$ (lower mass sources) \citep{peterson2011, paliya2024}, and the division between type A and type B quasars (more massive) at FWHM $\sim 4000$ km s$^{-1}$ \citep{marziani2003b, panda_marzena2024}.

The correlation between the FWHM and the accretion rate  confines the results to a well-defined region, with just a few outliers. The negative correlation between Eddington ratio and both FWHM and black hole mass is more apparent here, with low-mass black holes constrained to high accretion regimes. This pattern resembles the behavior of the optical Eigenvector 1 (EV1) sequence \citep{boroson1992, sulentic2000, marziani2003b, shen_ho_2014, panda2018, panda2019b}, which connects FWHM and optical Fe II strength, quantified by the ratio $R_{\rm FeII} = \text{EW(Fe,II}\lambda4570)/\text{EW(H}\beta)$. The observed linear relationship between $R_{\rm FeII}$ and the Eddington ratio \citep{zamfir2010} supports a comparison with our model predictions. However, observations do populate the region of low accretion rate and low FWHM, which is not seen in our model, thus confining its predictions to a more limited region.

The ratio of the FWHM to $\sigma$ in line profile analysis is commonly referred to as the shape (or form) factor. This ratio provides insight into the shape of the emission line and is often used to assess whether a profile resembles a Gaussian, Lorentzian, or another distribution. For a Gaussian, the shape factor is approximately 2.35; for a Lorentzian, it approaches zero. A rectangular profile yields a value of 3.46, a triangular one gives 2.45, and an edge-on rotating ring yields 2.83 \citep{collin2006}. Figure \ref{fig:line_shapefactor} shows how the shape factor varies with Eddington ratio, black hole mass, and viewing angle.

The results show that as the accretion rate increases, the shape factor decreases, particularly in higher-mass black holes. At low accretion rates, the shape factor remains relatively high, indicating more symmetric, Gaussian-like profiles. This suggests a more uniform velocity distribution in the BLR under these conditions. As the accretion rate increases, especially in massive systems, the shape factor decreases, indicating increasingly Lorentzian profiles with broader wings. This likely reflects enhanced radiation pressure and turbulence at high accretion rates, leading to more chaotic gas motion in the BLR.

This trend is consistent with EV1 findings, where sources with FWHM < 4000 km s$^{-1}$ typically show Lorentzian profiles and high accretion rates, while broader lines (FWHM > 4000 km s$^{-1}$) are associated with Gaussian shapes and lower accretion rates \citep{sulentic2000}. These observational trends agree with our predictions, including the role of high radiation pressure \citep{negrete2018}. The transition in shape factor does not occur at a fixed Eddington ratio, but also depends on black hole mass.

However, the model underestimates outflow velocities in low-mass systems ($\sim10^7 M_{\bullet}$). Even at high accretion rates, the line profiles remain strongly double-peaked due to insufficient vertical motion, keeping the BLR close to the disk and dominated by Keplerian dynamics. Nonetheless, the predicted BLR location remains consistent with observed trends and the lines are appropriately narrow.

Similarly, at a fixed accretion rate, the shape factor decreases with increasing black hole mass. This trend is stronger (and begins at lower black hole masses) for higher accretion rates. For low-mass black holes, the shape factor remains close to the Gaussian value, indicating a more symmetric velocity distribution. As the black hole mass increases, particularly at high accretion rates, the shape factor drops, indicating that line profiles become increasingly Lorentzian. This is likely due to broader wings caused by more extreme velocity fields.

While this may seem counterintuitive, for high black hole masses (e.g., $10^9 M_{\bullet}$, typical of quasars), the model predicts two distinct quasar populations: one exhibiting Gaussian line shapes and the other Lorentzian. The transition occurs near an accretion rate of $\dot{m} \sim 0.16$, which aligns with the empirical division between type A quasars (below the orange line in Fig.~\ref{fig:line_shapefactor}) and type B quasars (above the orange line, middle panel). For lower masses, the model predicts only Gaussian shapes, due to the insufficient outflow velocities discussed above.

The shape factor depends on viewing angle, showing a minimum at intermediate inclinations (between $20^\circ$ and $25^\circ$), especially for higher black hole masses. At low inclinations, it stays near 2.355, indicating a more symmetric BLR appearance and Gaussian-like line profiles when viewed face-on. Beyond $\sim 30^\circ$, the shape factor increases again, particularly for massive black holes, suggesting that edge-on views yield more Gaussian-like profile. This is possibly due to projection effects compressing the velocity field into a more symmetric form.

The intermediate-angle minimum likely reflects both projection effects and the complex BLR geometry in high-mass, high-Eddington systems. In these cases, the line of sight intersects outflow regions with enhanced velocity dispersion, producing more Lorentzian-like profiles.

\section{Time-delay and R-L relation}\label{sec:blrsize}

\subsection{Time delay extraction and R-L construction}
\label{sect:R_L_method}

From the grid of simulations, we obtained the 3D cloud distribution defining the BLR geometry based on the FRADO model. This distribution allows us to calculate how the central radiation propagates to the clouds and subsequently reaches the observer. The observed response depends on the 3D structure, cloud emissivity, and viewing angle. The reprocessing time within each cloud is neglected, as it is fast compared to light-travel delays across the BLR.

\begin{figure*}
        \centering
        \includegraphics[scale=0.4]{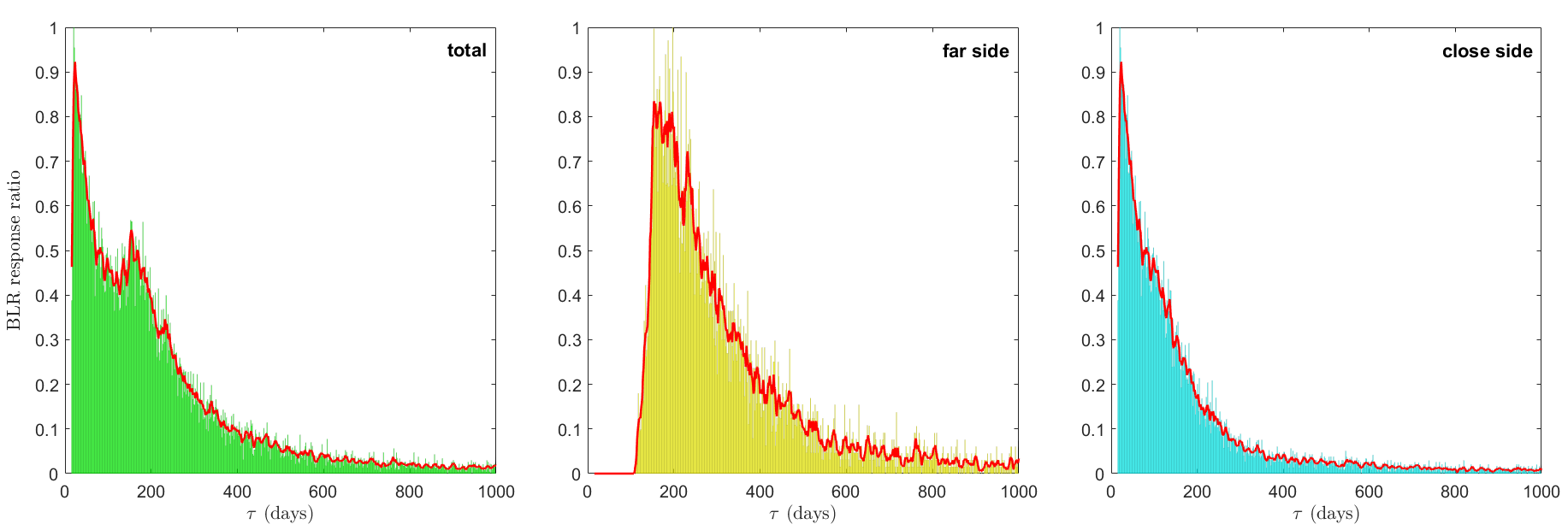}
        \caption{Time-delay histogram for a model corresponding to the mean quasar viewed at 39$^\circ$. }
        \label{fig:histogram}
\end{figure*}

A short pulse from the central source reaches the observer as an extended signal, spread over time by reprocessing in the BLR clouds. This response can be recorded as an emissivity-weighted time-delay histogram, representing the transfer function of the BLR. An example of such a transfer function is shown in the left panel of Figure~\ref{fig:histogram}. This case assumes full visibility of all clouds, without any shielding.

To explore obscuration effects, we divided the BLR into near and far sides. In cylindrical coordinates, the near side includes all clouds within the azimuthal range \( 0 \leq \phi < \pi \), corresponding to the vertical half of the BLR closest to the observer. The far side covers \( \pi \leq \phi < 2\pi \). In Cartesian coordinates, assuming the observer's line of sight lies in the x--z plane, clouds with \( x > 0 \) are considered to be on the near side; whereas those with \( x < 0 \) are on the far side. On this basis, we  constructed three time-delay histograms for each model: full, near-side only, and far-side only.

These histograms were analyzed to extract two key statistical quantities: the average time delay, $\tau_{\rm avg}$, and the peak time delay, $\tau_{\rm peak}$. The results are presented in Figure~\ref{fig:rlrelation}. The full histograms are typically asymmetric and often exhibit a double-peaked structure, with a sharper first peak originating from the near side and a broader, shallower second peak from the far side of the BLR. This decomposition makes the contribution from each side clearly visible.

\subsection{Results}

The theoretical R--L relation depends significantly on the assumptions described in Section~\ref{sect:R_L_method}. Strong asymmetry in the transfer function leads to a noticeable difference between the peak time delay and the average. Additionally, selective visibility of the near or far side can alter the estimated delay. Therefore, we explored various combinations and compare the results to observed trends. In practice, time delays are typically measured using standard methods without reconstructing the full transfer function, so it is not always clear which theoretical definition is most appropriate for comparison with observed R--L relations.

\subsubsection{Average time-delay option}

We first analyzed the R--L relation based on the average time delays. The results (shown in the upper-left panel of Figure~\ref{fig:rlrelation}) correspond to a fixed representative viewing angle of $39^\circ$, covering all considered black hole masses and Eddington ratios. The black hole mass is not shown explicitly, but for a fixed Eddington ratio, it can be inferred from the luminosity, whereby lower luminosities correspond to lower black hole masses.

The results deviate from observational trends, although the slope and overall location relative to the \citet{bentz2013} relation are promising. Specifically, the average time delays exhibit behavior opposite to that seen in observations with respect to accretion rate. The model predicts shorter delays for low Eddington ratio sources, consistent with \citet{bentz2013}, but longer delays for high accretion rates, in contrast to observational findings \citep[e.g.,][]{du2015}. This suggests that while the average delay reflects the overall BLR structure, it may not accurately capture the emission-line response seen in reverberation mapping, particularly under complex BLR dynamics.

We also explored potential biases that could arise from noise in observed BLR transfer functions. To test this, we removed simulated particles with response strengths below 10\% of the peak, aiming to exclude weakly responding regions that may contribute spurious delays. However, this adjustment did not resolve the discrepancy; specifically, the model’s prediction of increasing average time delays with higher accretion rates, which is not observed in reverberation data. This discrepancy suggests that the issue is likely intrinsically linked to the way the average delay reflects the BLR structure in the model.

\subsubsection{Peak time-delay option}

When focusing on the peak values of the time-delay histograms, we observed a strong consistency with the observed H$\beta$ R--L relation. The slope of the R-L relation obtained from our peak time delays aligns well with the empirical data, indicating that the FRADO model can accurately capture the distribution of material and the dynamic response of the BLR gas. Importantly, the spread in the predicted time delays, which is thought to be dependent to the accretion rate of the sources, also follows the trends observed in real AGNs; although we note that the observed spread is much larger than what we see in our simulations. This suggests that peak time delays can be the robust indicator of BLR geometry. 

In Figure~\ref{fig:rlrelation}, we plot the delays from the near and far sides of the distribution separately for each viewing angle. The difference between the far and near side time delays is increasing strongly with the viewing angle, which is expected. The near-side lags tend to fall below the standard R--L relation \citep{bentz2013} due to shorter light travel times, while far-side lags lie closer to or above it.

\subsubsection{R-L relation}

\begin{figure*}
        \centering
        \includegraphics[scale=0.44]{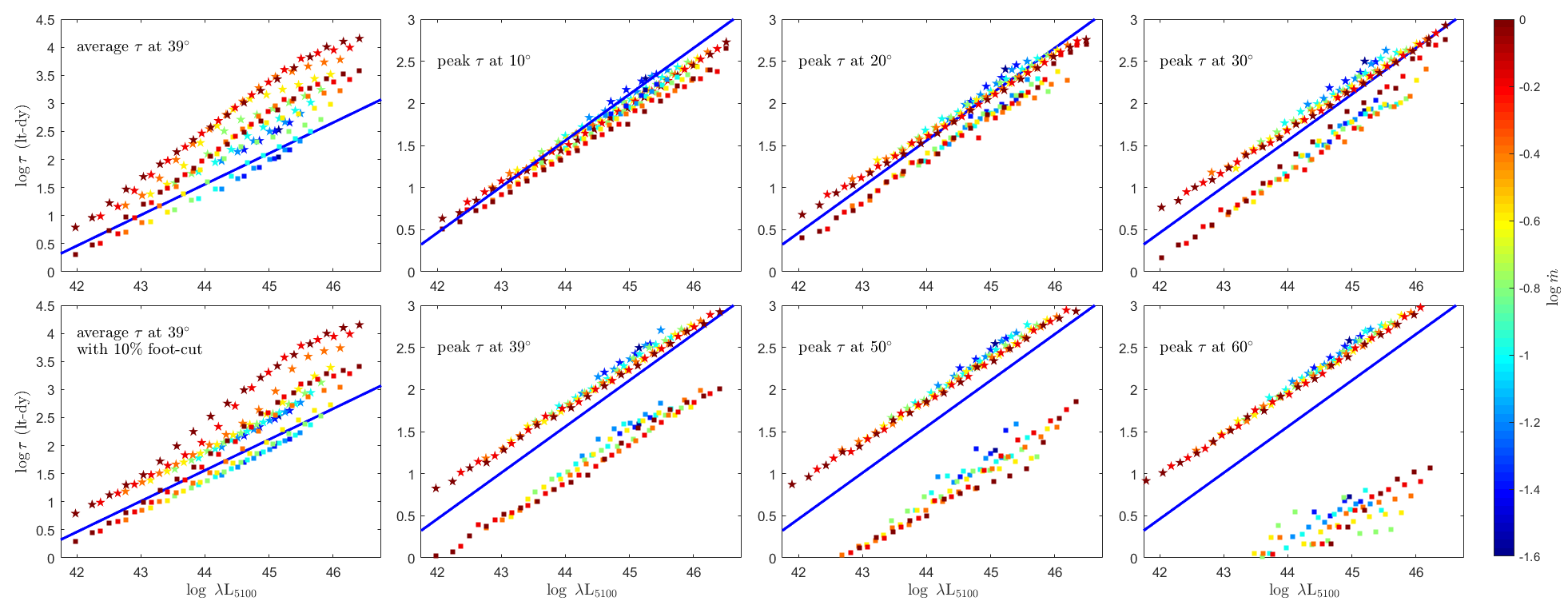}
        \caption{R–L relation for various viewing angles and BLR configurations, for the two options of measuring the time-lags: average values (left-most two panels) and peak values (other panels). The blue line in each panel represents the empirical R–L relation calibrated for low accretors from \citep{bentz2013}, the fully corrected fit with slope of 0.546 and interception value of 1.559. Color-coded data points correspond to different Eddington ratios as indicated by the colorbar. Pentagrams represent the far side of BLR relative to the observer, while squares correspond to the near side.}
        \label{fig:rlrelation}
\end{figure*}

The observed R--L relation should represent an average over all viewing angles. However, by examining the separate panels in Figure~\ref{fig:rlrelation}, we can draw the qualitative conclusion that the observed spread cannot be reproduced if we include all models uniformly. The much shorter time delays observed for high Eddington ratio sources can be accommodated in the model only when we selectively consider the near side of the BLR for such objects.

We note that this is not the same scenario as the one proposed by \citet{wang_shielding2014}, where the disk is assumed to shield the BLR near the equatorial plane and the outflow adopts a conical geometry, thereby shortening the delay from both near and far sides. In our case, we have assumed that there is some medium, such as the inner outflow, obscuring the far side of the BLR.

\section{Virial factor}

\begin{figure*}[htbp]
        \centering
        \includegraphics[scale=0.58]{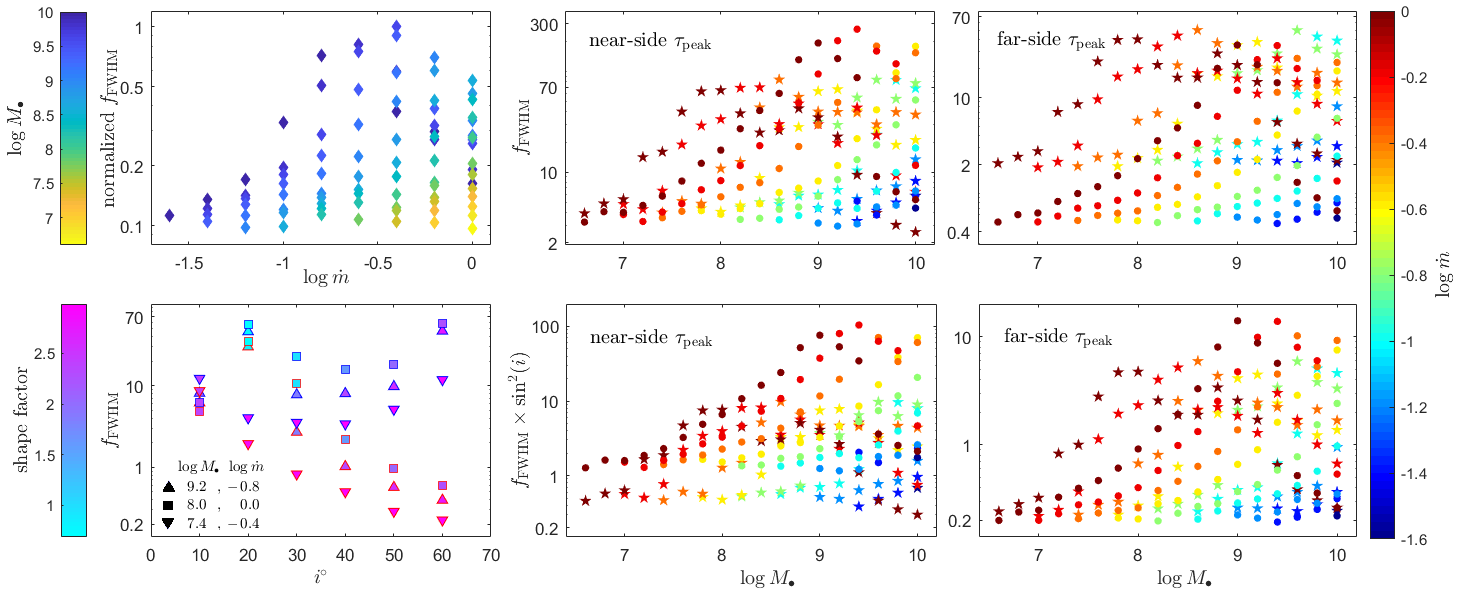}
        \caption{Dependence of $f_{\rm{fwhm}}$, predicted by 2.5D FRADO on Eddington ratio, black hole mass, viewing angle, and shape factor. Pentagrams and circles in four right panels represent the viewing angle of 20$^\circ$, and 39$^\circ$, respectively. In the top-left panel, the $f_{\rm{fwhm}}$ is shown vs. Eddington ratio for the far-side viewed at 60$^\circ$. In the bottom-left panel, the $f_{\rm{fwhm}}$ based on the far-side and near-side time-delays distinguished by edge colors of red and blue, respectively, is shown vs. viewing angle.}
        \label{fig:virial_factor_fwhm}
\end{figure*}

\begin{figure}[htbp]
        \centering
        \includegraphics[scale=0.6]{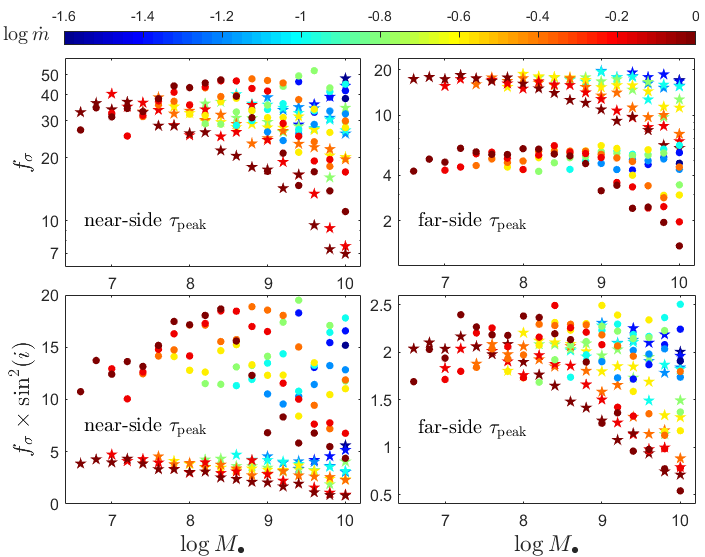}
        \caption{Dependence of $f_{\rm{\sigma}}$, predicted by 2.5D FRADO on Eddington ratio and black hole mass. Pentagrams and circles in four right panels represent the viewing angle of 20$^\circ$, and 39$^\circ$, respectively.}
        \label{fig:virial_factor_sigma}
\end{figure}

The virial factor (commonly denoted as $f$, also called correction factor) plays a critical role in estimating the mass of a supermassive black hole in AGNs using reverberation mapping techniques. Assuming a flattened disk-like BLR with predominantly Keplerian motion, the black hole mass, $ M_{\bullet}$, is typically estimated using the virial equation,
\begin{equation}
    M_{\bullet} = f ~ \frac{R_{\text{BLR}} }{G } ~ \Delta V ^2
,\end{equation}
where $R_{\text{BLR}}$ is the radius of the BLR (measured via reverberation mapping), $\Delta V$ is representative of velocity of particles in the BLR, and $ G $ is the gravitational constant. The virial factor accounts for uncertainties related to the geometry, kinematics, and inclination of the BLR in AGNs. Modeling of the BLR in 2.5D FRADO enables us to directly determine the virial factor (as discussed below). However, we will devote a separate comprehensive study on constraining the virial factor from our model in the near future.

\subsection{Virial factor based on the FWHM}

If we consider the observed FWHM of the broad emission line to be representative of velocity range of particles in the BLR, we can find $f$ via
\begin{equation}
 f_{\rm FWHM}   = \frac{G ~ M_{\bullet}}
    {c ~ \tau}
    \frac{1}{ \text{FWHM}^2_{\rm obs}}
, \end{equation}
where $\text{FWHM}_{\rm obs}$ is the observed FWHM and $i$ is the inclination angle (the angle between the observer's line of sight and the symmetry axis). The observed FWHM may be related to that of intrinsic through $\text{FWHM}_{\rm obs} \approx \text{FWHM}_{\rm int} \cdot \sin (i)$ \citep{collin2006}.

Our results for the dependence of $f_{\rm FWHM}$, as predicted by the 2.5D FRADO model, on the black hole mass, Eddington ratio, viewing angle, and shape factor are presented in Figure~\ref{fig:virial_factor_fwhm}. Obviously, the results imply a significant departure of the BLR from a virialized state. For low black hole masses and low Eddington ratios, the $f_{\rm FWHM}$ starts at smaller values and increases with both parameters. There is a peak at black hole masses around $\log M_{\bullet} \sim 9.5$ and Eddington ratios around $\log \dot{m} \sim -0.4$. For higher black hole masses, $f_{\rm FWHM}$ decreases again, particularly at higher accretion rates. This behavior suggests a strong dependence of $f_{\rm FWHM}$ on both black hole mass and Eddington ratio.

These results also indicate that $f_{\rm FWHM}$ decreases with increasing viewing angle across all three representative models, indicating a clear anti-correlation, especially for time delays measured from the far side of the BLR. The shape factor results similarly indicate that at higher inclinations, line profiles tend to be more Gaussian-like, while lower inclinations favor more Lorentzian-like shapes.

\subsection{Virial factor based on $\sigma_{\rm line}$}

Alternatively, we can also use the velocity dispersion of particles in BLR, $\sigma$, to compute the virial factor as
\begin{equation}
 f_{\sigma}  = \frac{G ~ M_{\bullet}}
    {c ~ \tau}
    \frac{1}{ {\sigma}^2_{\rm line}}
.\end{equation}

Figure \ref{fig:virial_factor_sigma} shows the dependence of $f_{\sigma}$ on the black hole mass. As we can see, the range for $f_{\sigma}~\times~ \sin^2{i}$ in bottom-right panel of the figure, spanning approximately from 0.5 to 2.5 (although with a noticeable scatter for high black hole masses) is very  consistent with the general observational constraints of \citet{onken2004, collin2006}. The inclination-independent results in the top-left panel, show the same trend as well; however, these values are ten times larger than the observational constraints. The other panels display a strong segregation, which is based on the viewing angle.


\section{Discussion} \label{sec:discussion}

Here, we address  the advantages and shortcomings of the FRADO model in detail. We start with the line profiles, described below.

\subsubsection{Black hole mass and line widths}
Higher mass black holes produce broader emission lines due to the higher orbital velocities of gas in the BLR. This is consistent with the virial theorem, which links the velocity dispersion of the BLR gas to the black hole mass \citep[e.g.,][]{peterson2004}. However, in high Eddington ratio sources, radiation pressure may also contribute significantly to the gas dynamics, potentially modifying the line widths and inferred masses \citep[e.g.,][]{marconi2008}.

\subsubsection{Accretion rate and line profile shape}
While the accretion rate is expected to influence the total flux of the line emission, its effect on the profile shape in FRADO is also significant. Higher accretion rates tend to result in more asymmetric profiles, possibly due to winds or other outflow phenomena \citep[e.g.,][]{netzer2013, elvis2000}. As the accretion rate increases, radiation pressure can dominate and reduce the width of the line profile, often introducing asymmetry. At lower accretion rates, the disk is colder, and the dust sublimation radius moves closer to the black hole, where Keplerian rotation is faster. This results in broader profiles due to higher orbital velocities in the BLR. This outcome is aligned with observational studies, whereby AGNs with high accretion rates (near or above the Eddington limit) tend to have narrower profiles, while low-accretion-rate AGNs display broader lines due to less interference from such forces \citep{pounds1995, DuPu2016}.

FRADO also predicts outflows and their importance rises with accretion rate, but the outflow appears to be underestimated for small black hole masses. This is evident in the double-peaked line profiles predicted by FRADO for low black hole masses and high accretion rates, in contrast to the observed single-peaked Lorentzian shapes in most Narrow Line Seyfert 1 galaxies.

The line width is consistent with observation, which suggests that the launching radius is properly evaluated; however, the outflow or turbulence is not strong enough to remove the double-peaked structure. We do not have a clear explanation for this. It may be that line driving plays a role once the clouds are already launched, but this extra force is only needed for low-mass objects in order to be consistent with observation. However, estimating the role of the line-driving force is beyond the scope of the current paper. This would also call for  the filtering of the continuum through the inner HIL region producing lines such as HeII~$\lambda1640$ and CIV~$\lambda1549$ to be modeled.
\subsubsection{Viewing angle and profile broadening}
The inclination of the BLR disk plays a crucial role in shaping the observed line profiles. Face-on views reveal the double-peaked structure expected from a rotating disk, while edge-on views produce broader and more asymmetric profiles. This behavior is consistent with models of disk-like BLR geometries \citep[e.g.,][]{gaskell2009}.

\subsection{Shape factor}

As presented in the results, with increasing accretion rate, the shape factor tends to decrease, especially for massive black holes, indicating a shift from more Gaussian-like to Lorentzian-like profiles. This may be attributed to an increase in turbulence and velocity dispersion in the BLR gas at higher accretion rates \citep[e.g.,][]{sulentic2000}. For higher black hole masses, the shape factor decreases further, particularly at high accretion rates. This suggests that the emission lines become broader and more Lorentzian-like for massive black holes, likely due to the stronger gravitational potential and more extreme velocity distributions in the BLR \citep[e.g.,][]{collin2006}. The shape factor also exhibits a minimum at intermediate viewing angles ($i \sim 25^{\circ}$), with Gaussian profiles being more prominent at both low and high inclinations. This behavior may be linked to projection effects and the complex geometry of the BLR, particularly with the formation of an outflow structure at higher black hole masses and Eddington ratios, where the line of sight intersects regions of the BLR with higher velocity dispersion, leading to Lorentzian-like profiles. The shape factor is less sensitive to detailed profile features than the presence or absence of a double-peak structure; therefore, the predicted trends remain consistent with the discussion in \citet{collin2006}.

\subsection{Time lags and the R-L relation}

The slope predicted by FRADO for high Eddington ratio sources is slightly flatter than the canonical slope of 0.5 expected from analytical FRADO \citep{czerny2011} and the result by \citet{bentz2013}. However, this is not surprising, taking into account newer findings for high Eddington ratio sources. The direct results from GRAVITY \citep{GRAVITY_4_objects_2024}, which resolved the BLR structure, show a shallower slope than \citet{bentz2013}. This is consistent with higher accretion rates leading to shorter time delays, although the uncertainties were large due to the small number of observed objects (four). Our model, in its current version, still underestimates the deviation of high-accretion sources relative to the standard relation of \citet{bentz2013}, meaning that for a given viewing angle and selected visibility (whether near side, far side, or both), the difference in time delays between high- and low-accretion sources is not comparable to the observed spread of $\sim$ 0.2 to 0.4 dex \citep{bentz2013, DuPu2016, du_wang2019, martinez2019}.

Although the model does not predict the spread directly, several mechanisms could contribute to the shortening of the time delay. At higher accretion rates, the inner disk region may become denser and more opaque. This could preferentially obscure the far side of the BLR, due to self-absorption or scattering \citep{netzer2010, wang_shielding2014}, or due to strong disk winds that rise above the BLR and block the far side \citep[e.g.,][]{murray1995, elvis2000}. Additionally, black hole spin affects the monochromatic flux and the ionization parameter in distinct ways, which could also influence the observed time delays \citep[e.g.,][]{czerny2019}.

The near side of the BLR contributes more strongly to the observed emission-line response, particularly at moderate to high inclinations, due to reduced obscuration and more favorable radiative conditions compared to the far side. Therefore, the systematic scatter observed in the R--L relation for high accretion-rate sources can be attributed to geometric asymmetries and radiative weighting of the line response across the BLR. Specifically, our modeling suggests that this deviation correlates with a progressive dominance of near-side emission in the reverberation signal. Although this effect may also apply to low accretors, as some are seen to deviate from the standard R--L relation of \citet{bentz2013}. At high inclinations, where the near side of the BLR is obscured by the dusty torus, the far side can remain visible in low-accretion systems due to the absence of inner disk winds. In contrast, in high-accretion systems, inner outflows also end up obscuring the far side. This might explain why some observed low accretors are found slightly above the standard R--L relation. Referring to the possible partial presence of matter-bounded clouds (as we have considered in this work), such conditions could indeed reduce the far-side emissivity, particularly in high-Eddington sources. Thus, this effects could contribute to the diminished response. However, this effect is expected to be minor; otherwise, we would expect the near-side emission to be similarly suppressed.

Comparing our results in Figure~\ref{fig:rlrelation} with the observational data \citep{martinez2019} indicates that the maximum viewing angle at which H$\beta$ is typically observed is around 39$^\circ$; beyond this angle, either the H$\beta$ flux becomes very weak or the transition to type 1.5 AGNs occurs. The results also imply that the standard R–L relation, calibrated primarily for low accretors \citep{bentz2013}, mostly includes sources observed at relatively low inclinations. In a forthcoming paper, we will present a more comprehensive analysis of the implications of this model for the R–L relation and its application to quasar cosmology.

These findings suggest that accurate black hole mass estimates using the virial method based on FWHM must account for multiple parameters, including the accretion rate, as it has a significant impact on the virial factor, particularly in the intermediate to high-mass regime.

On the other hand, the line dispersion, $\sigma$, is often considered a more reliable indicator of BLR kinematics than FWHM, as also supported by Figure~\ref{fig:virial_factor_sigma}. This is because $\sigma$ captures the full range of velocities within the BLR, including contributions from both random motions and the extended wings of the emission line profile. Studies such as that of \citet{peterson2004} have emphasized that $\sigma$ yields more accurate black hole mass estimates and reduces the scatter in virial factor calibrations. While the FWHM primarily reflects the velocity of the bulk of the gas, it can overlook high-velocity components, leading to potential underestimation of black hole masses. Moreover, the FWHM is more sensitive to inclination and projection effects. As shown in our results in Figure~\ref{fig:virial_factor_sigma}, using line dispersion may lead to reduced uncertainties in black hole mass determinations. For instance, \citet{collin2006} argued that adopting $\sigma$ instead of FWHM also decreases the scatter in the $M_{\bullet} - \sigma_{\star}$ relation, which connects black hole mass to the velocity dispersion of the host galaxy’s bulge.

There are important issues that are not addressed in the current version of the FRADO model. As mentioned earlier, the vertical velocity for small black hole masses appears to be underestimated and we suspect this may be related to the absence of the line-driving force. Our model launches clouds via dust radiation pressure, but it is possible that line driving also plays a role; although we currently have no estimate for its contribution. Combining the effects of both radiation pressure mechanisms poses a significant challenge. Existing codes for exclusively line-driven winds \citep{risaliti2010,Quera2023} are based on different frameworks and integrating them with our dust-driven model is not straightforward. Including the line-driving force could also impact our conclusions on the metallicity. In the current model, we adopted a metallicity that is five times higher than solar, as otherwise the launching force would become too weak.

Finally, we have not specified what occurs in the disk regions where the wind is not launched. For low-mass black holes, this situation arises (in our computations) at low Eddington ratios. In such cases, the disk surface may be directly irradiated by the central source, potentially giving rise to emission lines. On one hand, AGN disks are relatively flat and geometrically thin, so direct irradiation may not be efficient \citep[see e.g.,][]{loska2004}. However, irradiation could still occur via scattering in a highly ionized medium (e.g., a warm absorber) as also discussed in the cited work.

\subsection{FRADO and microlensing}\label{sec:microlensing}

Since the 2.5D FRADO BLR model is able to reproduce a wide variety of broad emission line profiles observed in AGNs, it is also of interest to test whether microlensing of the FRADO BLR can account for the line profile distortions seen in a number of gravitationally lensed AGN. A negative result would imply the need for a significant revision of the model.

Microlensing by stars in the lensing galaxy of a lensed AGN can selectively magnify subregions of the quasar, particularly within the BLR. This leads to observable distortions in the emission line profiles, which can be used to probe the size, geometry, and kinematics of the BLR. Based on generic BLR models, \citet{Hutsemekers2021}, \citet{Hutsemekers2023, Hutsemekers2024a, Hutsemekers2024b}, and \citet{Savic2024} have shown that Keplerian disk and equatorial wind models best describe the geometry and dynamics of the BLR; whereas polar wind models can be excluded. Using the physically motivated 2.5D FRADO model, with well-defined source parameters and avoiding arbitrary parameterization, we should be able to simultaneously reproduce both the intrinsic line profiles and their microlensing-induced distortions, thereby providing new constraints on the BLR structure and potentially direct estimates of physical parameters such as the black hole mass and accretion rate.

As a first step, we used the method described in \citet[][and references therein]{Hutsemekers2023, Hutsemekers2024a, Hutsemekers2024b}, replacing the generic BLR models with a 2.5D FRADO BLR model. We adopted a representative model with a black hole mass of $10^8 M_{\odot}$, an accretion rate of 0.1 in Eddington units, and an inclination of 34\degr. The isovelocity maps generated by FRADO were then convolved with microlensing magnification maps. For each position on the magnification map, a magnification profile, $\mu(v)$, was computed; this profile represents the wavelength-dependent distortion of the emission line profile caused by microlensing ( where $\mu(v) = 1$ corresponds to no distortion). The resulting profiles were then compared with a representative set of observed magnification profiles taken from \citet{Hutsemekers2023, Hutsemekers2024a, Hutsemekers2024b, Savic2024}, after scaling to a common velocity grid.

\begin{figure}
        \centering
        \includegraphics[scale=0.25]{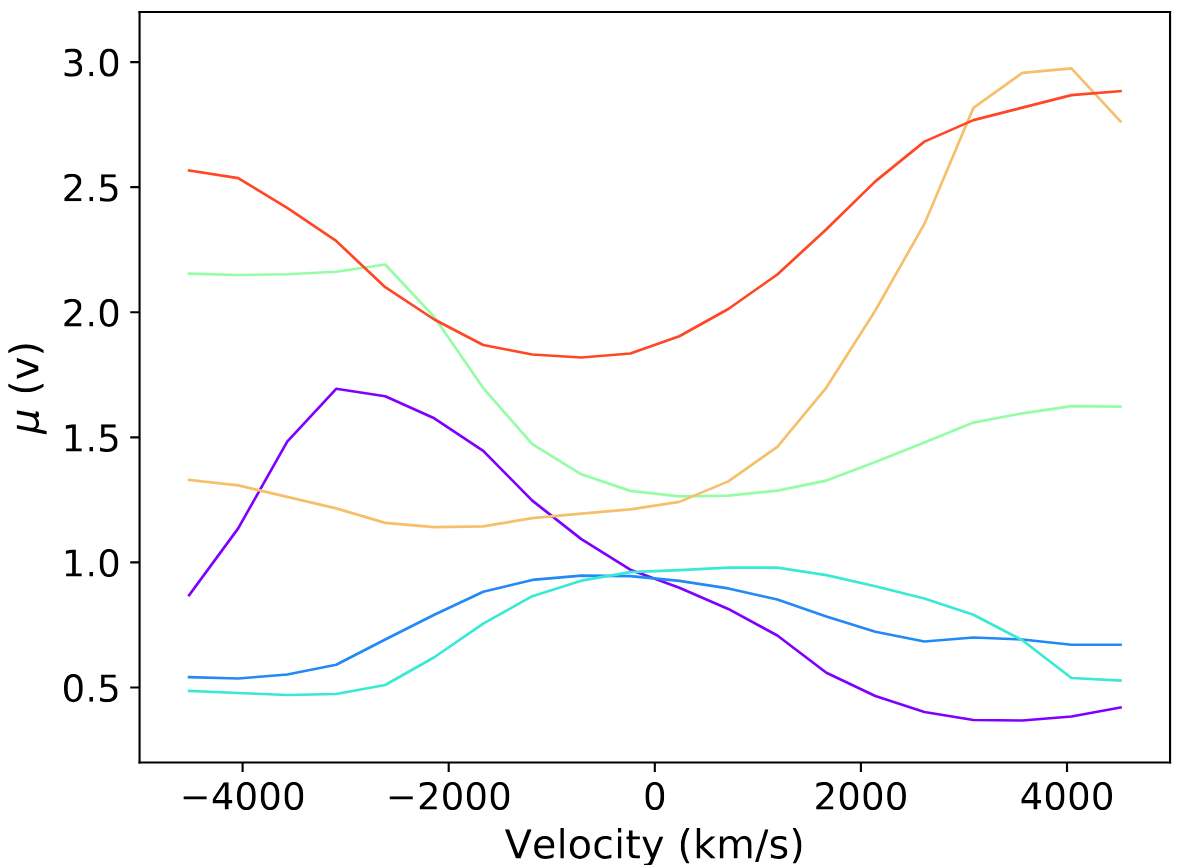}
        \caption{Magnification profiles generated by microlensing a 2.5D FRADO model of the BLR. These profiles have been selected to match six observational magnification profiles of different shapes, each one depicted by a randomly chosen color.}
        \label{fig:muv_frado}
\end{figure}

As illustrated in Fig.~\ref{fig:muv_frado}, microlensing of the 2.5D FRADO BLR model can reproduce the diverse line profile distortions observed in lensed AGNs. Both the blue and red wings of the lines can be magnified, demagnified, or affected asymmetrically. This result shows that by comparing a sufficiently large set of simulated line and magnification profiles to observed data, it could be possible to constrain BLR properties more effectively than with generic BLR models.

\section{Conclusion}\label{sec:conclusion}

Building on our previous work concerning the dynamical properties and structure of the low-ionization line (LIL) broad-line region (BLR) using the non-hydrodynamic single-cloud 2.5D FRADO model \citep{naddaf2021}, we tested the model by computing line profiles based on the predicted distribution of clouds along their trajectories for a relatively broad grid of initial black hole masses and accretion rates. The model is shown to accurately predict the location of the inner radius of the BLR where clouds are launched. The predicted line shape factor is broadly consistent with observational trends and we derived meaningful trends for the virial factor as a function of the model parameters. The line profiles are well reproduced, especially for larger black hole masses. The R-L relation is well captured, both in terms of slope and normalization, for sources with low Eddington ratios. At high Eddington ratios, the observed R-L trend can be qualitatively recovered if the selective obscuration of the far side of the BLR is assumed.

\begin{acknowledgements}
This project is supported by the University of Liege under Special Funds for Research, IPD-STEMA Program. DH is F.R.S.-FNRS Research Director. BC acknowledges the OPUS-LAP/GA ˇCR-LA bilateral project (2021/43/I/ST9/01352/OPUS 22 and GF23-04053L). MHN would like to thank Ashwani Pandey for the permission to use his part of CLOUDY results on dependence of H$\beta$ on photon flux. MLMA acknowledges financial support from Millenium Nucleus NCN2023${\_}$002 (TITANs) an ANID Millennium Science Initiative (AIM23-0001).
\end{acknowledgements}

\bibliographystyle{aa}
\bibliography{naddaf}

\end{document}